\documentclass[12pt]{article}
\usepackage{amssymb,amsmath,amsthm,amsfonts,amscd}
 \usepackage{graphicx}
\newcommand\be{\begin{equation}}
\newcommand\ee{\end{equation}}
\newcommand\bea{\begin{eqnarray}}
\newcommand\eea{\end{eqnarray}}
\newcommand\ket[1]{|#1\rangle}
\newcommand\bra[1]{\langle #1|}

\newcommand{\fatalpha}{{\bf \alpha \kern -0.44em \alpha}}
\newcommand{\fatsigma}{{\bf \sigma \kern -0.54em \sigma}}
\newcommand{\tpchi}{{\bf D \kern -0.35em D}}
\newcommand{\llambda}{{\bf \lambda \kern -0.45em \lambda}}

\renewcommand{\theequation}{\arabic{equation}}
\renewcommand{\theequation}{\thesection.\arabic{equation}}
\bibliography{plain}
\title{\bf Quantifying entanglement of two relativistic  particles via  decomposable optimal entanglement witnesses} \vspace{20mm}
\author{ M. A. Jafarizadeh$^{a,b,c}$
 \thanks{E-mail:jafarizadeh@tabrizu.ac.ir},
  M. Mahdian$^{a}$
 \thanks{E-mail:Mahdian@tabrizu.ac.ir}\\
$^a${\small Department of Theoretical Physics and Astrophysics,
University of Tabriz, Tabriz 51664, Iran.}  \\ $^b${\small Institute
for Studies in Theoretical Physics and Mathematics, Tehran
19395-1795, Iran.}\\$^c${\small Research Institute for Fundamental
Sciences, Tabriz 51664, Iran.}} \pagebreak

\vspace{20mm}

\begin{document}
\maketitle \vspace{15mm}
\newpage
\begin{abstract}

The study of ``Entanglement of Formation'' of a mixed state of a bipartite system in high-dimensional Hilbert space is not easy in general. So, we focus on determining the amount of entanglement for a bipartite mixed state based on the concept of  decomposable optimal entanglement witness (DOEW), that can be calculated as a minimum distance of an entangled state from the edge of positive partial transpose (PPT) states which has the most negative (positive) expectation value for non-PPT (bound) entangled states. We have constructed DOEWs based on the convex optimization method, then by using of it we quantify the degree of entanglement for two spin half particles under the Lorentz transformations. For convenience, we restrict ourselves
to 2D momentum subspace and under this constraint when the momentum and the Lorentz boost are parallel, we have shown that the entanglement is not relativistic invariant.


\end{abstract}

\section{Introduction}

In the recent years it became clear that quantum entanglement
\cite{Einstein} is one of the most important resources in the
rapidly growing field of quantum information processing because, quantum entangled states produce nonclassical
phenomena. Therefore, specifying that a particular quantum state is entangled
or not, must be important since for separable quantum states statistical properties can be explained entirely
by classical statistics.\\
If a density matrix, $\rho_{(A,B)}$, of a composite system (A,B) can be written as a sum of products of density matrices of its
components, $\rho_{(A)}$ and $\rho_{(B)}$ in the form  $ \rho_{(A,B)}=\sum_{i} q_{i} \ \rho^i_{(A)} \otimes  \rho^i_{(B)} ,\   \  0 \leqslant q_{i} \leqslant1$ and $\sum_{i}q_{i}=1$,  then the system is separable, otherwise it is \textit{entangled}.
The first and most widely used related criterion for distinguishing
entangled states from separable ones, is  PPT criterion, introduced by Peres \cite{peres}.
Furthermore, a necessary and sufficient condition for separability in $\mathcal{H}_2\otimes \mathcal{H}_2$ and $\mathcal{H}_2\otimes \mathcal{H}_3$ Hilbert spaces ( $ \mathcal{H}_d$ denotes the Hilbert space
with dimension $d$ endowed
with usual inner product denoted by $\langle\ .\ \rangle$) was shown by Horodecki \cite{horodecki1}, which was based on a previous work by
Woronowicz \cite{woro}. However, in higher
dimensions, there are PPT states that are nonetheless entangled,
as was first shown in Ref. \cite{horodecki2}, again based on Ref.
\cite{woro}. These states are called bound entangled states because
they have the peculiar property that no entanglement can be
distilled from them by local operations \cite{14}. So the PPT criterion
is not sufficient for separability. Another approach to distinguish separable states from entangled ones is entanglement witness (EW) \cite{terhal}. An EW for
a given entangled state $\rho$ is an observable $\mathcal{W}$,  whose
expectation value over all separable states is nonnegative, but
strictly negative on $\rho$. There is a correspondence between EWs
and linear positive (but not completely positive) maps via
Jamiolkowski isomorphism \cite{jamiolkowski}. As an example the
partial transposition is a positive map (PM). Despite of the fact that
EWs are designed mainly for detection of  entanglement, it
has been shown \cite{witmeas} that the optimal EW associated with
a density matrix $\rho$ -in a the sense that, the expectation value
of the optimal EW (associated with $\rho$) over $\rho$ is the most
negative value between the expectation values of other EWs over
$\rho$- can be used as measure of entanglement quantifying the
amount of entanglement of $\rho$.\par

Relativistic aspects of quantum
mechanics have recently attracted much attention, especially in the
context of the theory of quantum information and entanglement. Recently, several groups have focused their investigations on relativistic quantum entanglement \cite{Harshman,
Lee,Caban,Bartlett,Wigner,Adami,Terashimo,Czachor,Alsing,Ahn,Jafarizadeh,lamata,lamata1,Jian,Jason,Andr}. Peres \emph{et} al.\cite{peres1} have observed that the reduced spin density matrix of a single spin-$\frac{1}{2}$ particle
 is not a relativistic invariant, given that Wigner
rotations  entangle the spin with the particle momentum
distribution when observed in a moving frame \cite{wigner}. Gingrich and Adami
have shown that the entanglement between the spins of two
particles is carried over to the entanglement between the momenta of
the particles by the Wigner rotation, even though the entanglement
of the entire system is Lorentz invariant \cite{Adami}. Terashimo and Ueda
\cite{Terashimo} and Czarchor \cite{Czachor} suggested that the
degree of violation of the Bell inequality depends on the velocity
of the pair of spin-$\frac{1}{2}$ particles or the observer with
respect to the laboratory. Alsing and Milburn  studied
the Lorentz invariance of entanglement, and showed that the
entanglement fidelity of the bipartite state is preserved
explicitly. Instead of state vector in Hilbert space, they
have used a 4-component Dirac spinor or a polarization vector in
favor of quantum field theory \cite{Alsing}. Ahn  also calculated the
degree of violation of the Bell's inequality which decreases with
increasing of velocity of the observer \cite{Ahn}.
Most of the previous works
were concerned with  pure states although authors in
\cite{Jafarizadeh,lamata,lamata1} have considered mixed quantum states that are described by  superposition of momenta. In pervious works the measure of entanglement for pure states under the Lorentz transformation has been calculated using the Bell's inequality.
\par
In Ref.\cite{mah}, we investigated  spin-momentum correlation in single-particle and showed that  entanglement decreases with respect to the increasing of observer's velocity. However, in this work, we are concerned with degree of entanglement for two relativistic spin half  particles and will show that it is possible to use the idea of DOEW operators as measure of entanglement in high-dimensional Hilbert space because finding the region of separable states is not easy in general. Instead of
using a superposition of momenta for particles we use only two momentum
eigen states ($p_1$ and $p_2$) and restrict ourselves
to 2D momentum subspace and under this constraint,  our analysis will concentrate to indicate a new method for constructing DOEWs based on the convex optimization method which has been widely used in quantum information theory \cite{jaf1,jaf2,jaf3,jaf4}. Likewise, in Refs. \cite{hilbersh,16} a connection between Hilbert-Schmidt measure and optimal EW associated with a state has been discussed. However, we show that the constructed DOEW is  based on minimal distance of the corresponding entangled state from the edge of PPT states. So,
 the minimal distance or expectation value is most negative (positive) for non-PPT (bound) entangled states. At the end, by using the DOEW, we show that when momentum and Lorentz boost are parallel, the entanglement of the entire system is not invariant under the Lorentz transformation.\par
This paper is organized as follows:  Sec. II, is devoted
to two-particle relativistic quantum states. In Sec. III, we present the DOEWs which are constructed using the convex optimization method.
In Sec. IV, we derive a simple formula for optimality of the decomposable EW via Hilbert-Schmidt measure. In Sec. V, we explicitly calculate
 the amount of entanglement of two relativistic particles.
The last section contains concluding remarks.

\section{ Two relativistic particles  quantum states}

Suppose we have a bipartite system with its quantum degrees of
freedom distributed among two parties $\mathcal{A}$ and
$\mathcal{B}$ with Hilbert spaces $\mathcal{H}_A$ and $\mathcal{H}_B$, respectively. The quantum states of one particle is made by two degrees of freedom: momentum \emph{p} and spin.
The former is a continuous variable with Hilbert space of infinite
dimension but in this work, we have only two eigen states $p_1$ and $p_2$, while the latter is a discrete one with Hilbert space of spin particle. The pure quantum state of such a system can always be written as
\be\label{p1}\ket{\psi}=\sum_{i=1}^{2}\sum_{j=1/2}^{-1/2}
c_{ij}\ket{p_i}\otimes \ket{j},\ee\\
where both $\ket{p_{1}}$ and $\ket{p_{2}}$ are two momentum eigen states of each particle and the kets $\ket{\frac{1}{2}}$ and $\ket{\frac{-1}{2}}$ are the eigenvectors of spin operator $\sigma_z$. $c_{ij}$'s are complex coefficients such that $\sum_{i,j}|c_{ij}|^2=1.$ Using the general pure state (\ref{p1}), we define  four orthogonal maximal entangled Bell states for one-particle in terms of momentum and spin states as  follows,

$$ \ket{\psi_1}=\frac{1}{\sqrt{2}}(\ket{p_{1}}\otimes\ket{\frac{1}{2}}+\ket{p_{2}}\otimes\ket{-\frac{1}{2}}) ,$$
$$ \ket{\psi_2}=\frac{1}{\sqrt{2}}(\ket{p_{1}}\otimes\ket{\frac{1}{2}}-\ket{p_{2}}\otimes\ket{-\frac{1}{2}}) ,$$
$$ \ket{\psi_3}=\frac{1}{\sqrt{2}}(\ket{p_{2}}\otimes\ket{\frac{1}{2}}+\ket{p_{1}}\otimes\ket{-\frac{1}{2}}) ,$$
\be\label{pure2}\ket{\psi_4}=\frac{1}{\sqrt{2}}(\ket{p_{2}}\otimes\ket{\frac{1}{2}}-\ket{p_{1}}\otimes\ket{-\frac{1}{2}}).\ee\\
Then, using the upper Bell states, we construct two-particle quantum states as
$$\ket{\psi_{\pm}}^{(1,2)}=\frac{1}{\sqrt{2}}(|\psi_1\rangle|\psi_1\rangle\pm|\psi_2\rangle|\psi_2\rangle),\quad
\ket{\phi_{\pm}}^{(1,2)}=\frac{1}{\sqrt{2}}(|\psi_1\rangle|\psi_2\rangle\pm|\psi_2\rangle|\psi_1\rangle),$$
$$\ket{\psi_{\pm}}^{(3,4)}=\frac{1}{\sqrt{2}}(|\psi_3\rangle|\psi_3\rangle\pm|\psi_4\rangle|\psi_4\rangle),\quad
\ket{\phi_{\pm}}^{(3,4)}=\frac{1}{\sqrt{2}}(|\psi_3\rangle|\psi_4\rangle\pm|\psi_4\rangle|\psi_3\rangle),$$
$$\ket{\psi_{\pm}}^{(1,3)}=\frac{1}{\sqrt{2}}(|\psi_1\rangle|\psi_1\rangle\pm|\psi_3\rangle|\psi_3\rangle),\quad
\ket{\phi_{\pm}}^{(1,3)}=\frac{1}{\sqrt{2}}(|\psi_1\rangle|\psi_3\rangle\pm|\psi_3\rangle|\psi_1\rangle),$$
$$\ket{\psi_{\pm}}^{(2,4)}=\frac{1}{\sqrt{2}}(|\psi_2\rangle|\psi_2\rangle\pm|\psi_4\rangle|\psi_4\rangle),\quad
\ket{\phi_{\pm}}^{(2,4)}=\frac{1}{\sqrt{2}}(|\psi_2\rangle|\psi_4\rangle\pm|\psi_4\rangle|\psi_2\rangle),$$
$$\ket{\psi_{\pm}}^{(1,4)}=\frac{1}{\sqrt{2}}(|\psi_1\rangle|\psi_1\rangle\pm|\psi_4\rangle|\psi_4\rangle),\quad
\ket{\phi_{\pm}}^{(1,4)}=\frac{1}{\sqrt{2}}(|\psi_1\rangle|\psi_4\rangle\pm|\psi_4\rangle|\psi_1\rangle),$$
\begin{equation}\label{bell0}
\hspace{1.7cm}\ket{\psi_{\pm}}^{(2,3)}=\frac{1}{\sqrt{2}}(|\psi_2\rangle|\psi_2\rangle\pm|\psi_3\rangle|\psi_3\rangle),\quad
\ket{\phi_{\pm}}^{(2,3)}=\frac{1}{\sqrt{2}}(|\psi_2\rangle|\psi_3\rangle\pm|\psi_3\rangle|\psi_2\rangle),
\end{equation}

and introduce the following orthonormal entangled states in 16-dimensional Hilbert space:

$$\ket{\Phi^1}=\cos\theta\ket{\psi_{+}}^{(1,2)}+\sin\theta\ket{\psi_{+}}^{(3,4)},\quad
\ket{\Phi^2}=\cos\theta\ket{\psi_{-}}^{(1,2)}+\sin\theta\ket{\psi_{-}}^{(3,4)},$$
$$\ket{\Phi^3}=\cos\theta\ket{\psi_{+}}^{(3,4)}-\sin\theta\ket{\psi_{+}}^{(1,2)},\quad
\ket{\Phi^{4}}=\cos\theta\ket{\psi_{-}}^{(3,4)}-\sin\theta\ket{\psi_{-}}^{(1,2)},$$
$$\hspace{0.1cm}\ket{\Phi^5}=\cos\theta\ket{\phi_{+}}^{(1,2)}+\sin\theta\ket{\phi_{+}}^{(3,4)},\quad
\ket{\Phi^{6}}=\cos\theta\ket{\phi_{-}}^{(1,2)}+\sin\theta\ket{\phi_{-}}^{(3,4)},$$
$$\ket{\Phi^7}=\cos\theta\ket{\phi_{+}}^{(3,4)}-\sin\theta\ket{\phi_{+}}^{(1,2)},\quad
\ket{\Phi^{8}}=\cos\theta\ket{\phi_{-}}^{(3,4)}-\sin\theta\ket{\phi_{-}}^{(1,2)},$$
$$\ket{\Phi^{9}}=\cos\theta\ket{\phi_{+}}^{(2,4)}-\sin\theta\ket{\phi_{+}}^{(1,3)},\quad
\ket{\Phi^{10}}=\cos\theta\ket{\phi_{+}}^{(1,3)}+\sin\theta\ket{\phi_{+}}^{(2,4)},$$
$$\ket{\Phi^{11}}=\cos\theta\ket{\phi_{-}}^{(2,4)}-\sin\theta\ket{\phi_{-}}^{(1,3)},\quad
\ket{\Phi^{12}}=\cos\theta\ket{\phi_{-}}^{(1,3)}+\sin\theta\ket{\phi_{-}}^{(2,4)},$$
$$\ket{\Phi^{13}}=\cos\theta\ket{\phi_{+}}^{(2,3)}-\sin\theta\ket{\phi_{+}}^{(1,4)},\quad
\ket{\Phi^{14}}=\cos\theta\ket{\phi_{+}}^{(1,4)}+\sin\theta\ket{\phi_{+}}^{(2,3)},$$
\begin{equation}\label{iso}\hspace{1.6cm}\ket{\Phi^{15}}=\cos\theta\ket{\phi_{-}}^{(2,3)}-\sin\theta\ket{\phi_{-}}^{(1,4)},\quad
\ket{\Phi^{16}}=\cos\theta\ket{\phi_{-}}^{(1,4)}+\sin\theta\ket{\phi_{-}}^{(2,3)}.
\end{equation}

In this work, we choose $\theta=\frac{\pi}{4}$ in the pure states (\ref{iso}), then we obtain the so-called Bell-type states which are
maximally entangled states. Now, based on momentum distribution we have two types of pure states as the following: \\
{\bf The first type of pure states}\\
 The first type is $\ket{\Phi^i}$ with $i=$ add number, this type of pure states are made by $\sum_{i,j=1}^2\pm \ket{p_i,j}\otimes \ket{p_i,j}$ or $\sum_{\{i\ne j\}=1}^2 \pm\ket{p_i,i}\otimes \ket{p_i,j}$, which have the same momentum $p_i$ for two particles.\\
{\bf The second  type of pure states}\\
The second  type is $\ket{\Phi^i}$ with $i=$ even number, this pure states are made by  $\sum_{i,j=1}^2 \pm \ket{p_i,i}\otimes \ket{p_j,j}$ or $\sum_{i,j=1}^2 \pm \ket{p_i,i}\otimes \ket{p_j,i}$, which have different momentum for each particle.\\
{\bf Two types of mixed density matrices}\\
Consider a two-particle  quantum mixed state which is defined as a convex combination of the pure states (\ref{iso}), i.e. , \be\label{ro1}\rho=\sum_{i=1}^{16} q_i
\ket{\Phi^{i}} \bra{\Phi^{i}},\ee where $q_i\geq 0$, $\sum _{i}^{16} q_i=1 .$
According to the definition of two types of pure states, we construct two types of  mixed density matrices as
\begin{equation}\label{de1}
\quad \rho_1=\Sigma_{i=1}^{16}\emph{q}_{i}
\ket{\Phi^i}\bra{\Phi^i}  \ ,\ \ 0\leq \emph{q}_{i} \leq 1 \ , \
\sum_{i=1}^{16}\emph{q}_{i}=1,  \ \mathrm{\emph{i} = odd \ number,}
\end{equation}
\begin{equation}\label{de2}
\quad \rho_2=\Sigma_{j=1}^{16}\emph{q}_{j}
\ket{\Phi^j}\bra{\Phi^j}  \ ,\ \ 0\leq \emph{q}_{j} \leq 1 \ , \
\sum_{j=1}^{16}\emph{q}_{j}=1,  \ \mathrm{\emph{j} = even \ number.}
\end{equation}
{\bf Relativistic quantum states}\\
In this article, we assume that spin is in the z-direction and momentums are in the yz-plane, i.e., ${p_{1(2)}}=(0,
p_{1(2)}\sin{\theta_{1(2)}}, p_{1(2)}\cos{\theta_{1(2)}}) $ and for an observer in another reference frame $S$ described by
an arbitrary Lorentz boost $\Lambda$, the transformed pure states (see (\ref{wig2}) in Appendix A )
are given by
\begin{equation}\label{pure1}\ket{\Phi^{i}}\longrightarrow U(\Lambda_1)\otimes U(\Lambda_2)\ket{\Phi^{i}} .
\end{equation}
For example
$$\hspace{-5mm}\ket{\Lambda \Phi^{1}}=U(\Lambda_1)\otimes U(\Lambda_2)\ket{\Phi^1}=\ket{\Lambda p_1,n_1}\ket{\Lambda p_1,n_1}+\ket{\Lambda p_2,n^\prime_2}\ket{\Lambda p_2,n^\prime_2}+\ket{\Lambda p_2,n_2}\ket{\Lambda p_2,n_2}+\ket{\Lambda p_1,n^\prime_1}\ket{\Lambda p_1,n^\prime_1},$$
where
$$n_{1(2)}=D(W,p_{1(2)})\ket{0} ,\ \ n^\prime_{1(2)}=D(W,p_{1(2)})\ket{1}.$$\\ The kets  $\ket{\Lambda p_{1}}$ and $\ket{\Lambda p_{2}}$ are two orthogonal momentum eigen-state after the Lorentz transformation. Thus, we made two types of mixed density matrices ($\rho_1$ and $\rho_2$) and after some calculation we can see that the mixed state  $\rho_2$ doesn't change under the Lorentz transformation. But, in spacial case, when the momentum and the boost direction are parallel then the mixed density matrix  $\rho_1$ which was constructed by the first type of pure states, changes under the Lorentz transformation that we will consider in the next section. Let us first obtain the feasible region (FR) of the relativistic density matrix $\rho_1$ as follows:\\
{\bf Feasible region of the relativistic mixed density matrix $\rho_1$}

We consider the density matrix of
(\ref{de1}) and by imposing PPT conditions with respect to each parties, we
obtain the FR. For this particular density
matrix, the positivity of partial transpositions gives linear
constraints on the parameters $q_i$ (where \emph{i} is odd number).
The PPT condition with respect to the first party implies that the eigenvalues of $\rho^T_1$ given by

$$\hspace{-10mm}\lambda_{1(2)}=\frac{16(q_{10}+q_{14})\cos^4{\frac{\theta_{1(2)}}{2}}}{\cos^4{\frac{\theta_{1}}{2}}+\cos^4{\frac{\theta_{2}}{2}}},\quad \hspace{10mm}\lambda_{3(4)}=\frac{16(q_{12}+q_{16})\cos^4{\frac{\theta_{1(2)}}{2}}}{\cos^4{\frac{\theta_{1}}{2}}+\cos^4{\frac{\theta_{2}}{2}}},$$
$$\hspace{-13mm}\lambda_{5(6)}=\frac{16(q_{2}+q_{5})\cos^4{\frac{\theta_{1(2)}}{2}}}{\cos^4{\frac{\theta_{1}}{2}}+\cos^4{\frac{\theta_{2}}{2}}},\quad
\hspace{14mm}\lambda_{7(8)}=\frac{16(q_{1}+q_{6})\cos^4{\frac{\theta_{1(2)}}{2}}}{\cos^4{\frac{\theta_{1}}{2}}+\cos^4{\frac{\theta_{2}}{2}}},$$
$$\hspace{-4mm}\lambda_9=\frac{16(q_{9}-q_{13})\cos^2{\frac{\theta_{2}}{2}}\cos^2{\frac{\theta_{1}}{2}}}{\cos^4{\frac{\theta_{1}}{2}}+\cos^4{\frac{\theta_{2}}{2}}},\quad
\hspace{8mm}\lambda_{10}=\frac{16(-q_{9}+q_{13})\cos^2{\frac{\theta_{2}}{2}}\cos^2{\frac{\theta_{1}}{2}}}{\cos^4{\frac{\theta_{1}}{2}}+\cos^4{\frac{\theta_{2}}{2}}},$$
$$\hspace{-3mm}\lambda_{11}=\frac{16(q_{11}-q_{15})\cos^2{\frac{\theta_{2}}{2}}\cos^2{\frac{\theta_{1}}{2}}}{\cos^4{\frac{\theta_{1}}{2}}+\cos^4{\frac{\theta_{2}}{2}}},\quad
\hspace{6mm}\lambda_{12}=\frac{16(-q_{11}+q_{15})\cos^2{\frac{\theta_{2}}{2}}\cos^2{\frac{\theta_{1}}{2}}}{\cos^4{\frac{\theta_{1}}{2}}+\cos^4{\frac{\theta_{2}}{2}}},$$
$$\hspace{-7mm}\lambda_{13}=\frac{16(q_{3}-q_{5})\cos^2{\frac{\theta_{2}}{2}}\cos^2{\frac{\theta_{1}}{2}}}{\cos^4{\frac{\theta_{1}}{2}}+\cos^4{\frac{\theta_{2}}{2}}},\quad
\hspace{10mm}\lambda_{14}=\frac{16(-q_{3}+q_{5})\cos^2{\frac{\theta_{2}}{2}}\cos^2{\frac{\theta_{1}}{2}}}{\cos^4{\frac{\theta_{1}}{2}}+\cos^4{\frac{\theta_{2}}{2}}},$$
$$\hspace{-7mm}\lambda_{15}=\frac{16(q_{1}-q_{7})\cos^2{\frac{\theta_{2}}{2}}\cos^2{\frac{\theta_{1}}{2}}}{\cos^4{\frac{\theta_{1}}{2}}+\cos^4{\frac{\theta_{2}}{2}}},\quad
\hspace{10mm}\lambda_{16}=\frac{16(-q_{1}+q_{7})\cos^2{\frac{\theta_{2}}{2}}\cos^2{\frac{\theta_{1}}{2}}}{\cos^4{\frac{\theta_{1}}{2}}+\cos^4{\frac{\theta_{2}}{2}}},$$
$$
$$
must be nonnegative, i.e., we have the constraints  $\lambda_{i}\geq0$ , for  $i=1...16$ which give
$$ q_9=q_{13}, \ \ q_{11}=q_{15}, \ \ q_3=q_5, \ \ q_1=q_7,$$
and
$$q_1+q_3+q_{11}+q_9=\frac{1}{2},$$
where we have used the normalization condition $\sum_{i} \emph{q}_{i}=1$.\\
similarly, by imposing the PPT condition to the second party, we obtain
\begin{equation}\label{fr1}
(\frac{1}{4}-q_i)\geq0,
\end{equation}
It is important to see that the FR of the density matrix $\rho_1$ is independent of the Lorentz transformation.

\section{Construction of DOEWs via convex optimization}
We have seen how to manipulate the relativistic density matrix  $\rho_1$ that we want to quantify its amount of entanglement, by using DOEW. So the next section deals with the basic definition of convex optimization and our scheme to construct DOEW by an exact convex optimization method which can be generalized for other density matrices.
\subsection{Entanglement witnesses}
An entanglement witness acting on the Hilbert space
${\cal{H}}={\cal{H}}_{4}\otimes {\cal{H}}_{4}$
is a Hermitian operator $\mathcal{W}=\mathcal{W}^{\dag}$, that satisfies
$Tr(\mathcal{W}\rho_s)\geqslant0$ for any separable state $\rho_s$ in
$ {\textbf{B}}({\cal{H}})$ (Hilbert space of bounded
operators), and has at least one negative eigenvalue. If a density
matrix $\rho$ satisfies $Tr(\mathcal{W}\rho)<0$, then $\rho$ is an
entangled state and we say that $\mathcal{W}$ detects entanglement of
the density matrix $\rho$. The existence of an EW for any entangled state is a direct consequence of Hahn-Banach theorem \cite{Hahn} and the fact that the space of separable density operators is convex and closed. Geometrically,
EWs can be viewed as hyper planes that separate some
entangled states from the set of separable states and, hyper plane
indicated as a line corresponds to the state with $Tr[\mathcal{W}\rho]=0$.

\textbf{Definition 1.}
An EW $\mathcal{W}$ is said to be optimal,
if for all positive operators $\mathcal{P}$ and $\varepsilon>0$, the new
Hermitian operator
\begin{equation}\label{Wn2}
\mathcal{W'}=(1+\varepsilon) \mathcal{W}-\varepsilon \mathcal{P}
\end{equation}
is not anymore an EW \cite{Lewenstein1}.
Suppose that there is a positive operator $\mathcal{P}$ and $\epsilon\geq0$
such that  $\mathcal{W}_{new}=\; \mathcal{W}_{opt.}-\epsilon \mathcal{P} $ is yet an EW. This
means that if $Tr(\mathcal{W}_{opt.}\rho_s)=0$, then $Tr(\mathcal{P}\rho_s)=0$, for
all separable states $\rho_s$. Also, one can assume that the positive
operator $\mathcal{P}$ is a pure projection operator, since an arbitrary
positive operator can be written as convex combination of pure
projection operators with positive coefficients.

\textbf{Definition 2.}
When talking about EW's one has to distinguish two different
kinds. On the one hand, there are the so-called decomposable
EWs (DEW), which can be written as
\begin{equation}\label{decom}
\mathcal{W}=\mathcal{P}+\mathcal{Q}_{1}^{T_{A}},
 \quad\quad \mathcal{P},\mathcal{Q}_{1}\geq0,
\end{equation}
where the operator $\mathcal{Q}_{1}$ is positive semidefinite. It can
be easily verified that such witnesses cannot detect any
bound entangled states. $\mathcal{W}$ is non-decomposable EW if it can not be
put in the form (\ref{decom}) (for more details see
\cite{Doherty3}).
One should notice that only non-decomposable EWs can
detect PPT entangled states.
\par
\subsection{Construction of DOEWs via convex optimization }

At first, we expand an EW for a bipartite system ${\cal{H}}_{4}\otimes {\cal{H}}_{4}$ as follows
\begin{equation}\label{bpco1}
   \mathcal{W}=I_{4}\otimes I_{4}+\sum_{i,j=1}^{4}\mathcal{A}_{i,j} \hat{Q_{i}}\otimes \hat{Q'_{j}}
\end{equation}
where $I_{4}$ is a $4\times 4$ identity matrix, $\mathcal{A}_{i,j}$ are the parameters whose
ranges must be determined such that $\mathcal{W}$ be DOEW, and $Q_{i}$s ( $Q'_{j}$s ) are  Hermitian operators from the first ( second ) party of the Hilbert space. Via the mapping
\begin{equation}\label{Mapping}
P_{i} = Tr(Q_{i} \rho_{s}), \ \ P'_{j} = Tr(Q'_{j} \rho_{s})
\end{equation}
the set of separable states can be viewed as a convex region called FR.\\
Let $\{Q^i\}$ be bases for the space of Hermitian traceless matrices that operate
on 4-dimensional Hilbert space that can be written in terms of square matrices $E_{_{ij}}$, which have the value 1 at the position $(i , j )$ and zeros elsewhere \cite{Pf}, as
$$Q^1_{S}=\frac{1}{\sqrt{2}}( E_{_{1,2}}+E_{_{2,1}} ),\quad
Q^2_{S}=\frac{1}{\sqrt{2}}( E_{_{1,3}}+E_{_{3,1}} ),\quad
Q^3_{S}=\frac{1}{\sqrt{2}}( E_{_{1,4}}+E_{_{4,1}} ),$$
$$Q^4_{S}=\frac{1}{\sqrt{2}}( E_{_{2,3}}+E_{_{3,2}} ),\quad
Q^5_{S}=\frac{1}{\sqrt{2}}( E_{_{2,4}}+E_{_{4,2}} ),\quad
Q^6_{S}=\frac{1}{\sqrt{2}}( E_{_{3,4}}+E_{_{4,3}} ),$$
$$Q^7_{A}=\frac{i}{\sqrt{2}}( E_{_{2,1}}-E_{_{1,2}} ),\quad
Q^8_{A}=\frac{i}{\sqrt{2}}( E_{_{4,1}}-E_{_{1,4}} ),\quad
Q^9_{A}=\frac{i}{\sqrt{2}}( E_{_{4,2}}-E_{_{2,4}} ),$$
$$\hspace{3mm}Q^{10}_{A}=\frac{i}{\sqrt{2}}( E_{_{3,1}}-E_{_{1,3}} ),\quad
Q^{11}_{A}=\frac{i}{\sqrt{2}}( E_{_{3,2}}-E_{_{2,3}} ),\quad
Q^{12}_{A}=\frac{i}{\sqrt{2}}( E_{_{4,3}}-E_{_{3,4}} ),$$
\be\label{mat1}
Q^{13}=E_{1,1},\quad Q^{14}=E_{2,2},\quad Q^{15}=E_{3,3},\quad Q^{16}=E_{4,4}.\ee
where the subscript S and A indicates symmetric and antisymmetric operators, respectively. Notice that for $E_{i,j}=|i\rangle\langle j|$ where $i,j=1,...,4,$ the FR becomes
\begin{equation}\label{set3}
    \sum_{i,j=1}^{4} |P_{i,j}|^{2} \leqslant1.
\end{equation}\par
As mentioned above, we will use two steps towards the finding the parameters $\mathcal{A}_{i,j}$ for the density matrix $\rho_1$ and  fully characterize DOEWs based on exact convex optimization method for two partite system (see Appendix B).

{\bf The first step}

In the first step, according to the Hermitian traceless basis (\ref{mat1}), we introduced the maps (\ref{Mapping}) for any separable state $\rho_{s}$ which map the convex set of separable states to a bounded convex region that will be named FR. Then, recalling the definition of an EW, we impose the first condition which is the problem of the minimization of expectation
values of witness operators $\mathcal{W}$ with respect to separable states, i.e.,
$$min \ \ Tr(\mathcal{W} \rho_s)\geq0,$$
where $Tr(\mathcal{W} \rho_{s})$ is the objective function and the inside of the FR will be the inequality constraints. So, using the EW of (\ref{bpco1}) after some calculations one arrives at
\begin{equation}\label{op1}
   Tr(\mathcal{W}\rho_{s}) = 1+\sum_{i,j=1}^4 \mathcal{A}_{i,j} P_{i}P'_{j}\geqslant 0,
\end{equation}
which must be satisfied for all $P_i$ and $P^\prime_j$  belonging to the feasible region. In order to satisfy this condition, it is sufficient that the minimum value of $Tr(\mathcal{W}\rho_{s})$ be non-negative. Using standard convex optimization, we find this minimum value then we impose the non-negativity condition on this minimum. As mentioned before, for the bipartite system the FRs are $ \sum_{i=1}^{16} P_{i}^{2}\leqslant1 $ and $ \sum_{j=1}^{16} P_{j}'^{2}\leqslant1 $ which are in matrix notation as $P^{t} P\leqslant 1$ and $P'^{t} P'\leqslant 1.$ Then the Eq.(\ref{op1}) can be written as a convex optimization problem in the form
$$
\mathrm{Minimize} \quad 1 + P^{t}\mathcal{A} P',
$$
$$
\mathrm{subject \ to} \quad P^{t} P\leqslant 1,
$$
$$
\hspace{19mm}\quad P'^{t} P'\leqslant 1.
$$
We now write the primal Lagrangian of this problem, which will be helpful in the following development,
\begin{equation}\label{Lag1}
    L=1 + P^{t}\mathcal{A}P'-\frac{1}{2}\lambda (P^{t} P -1)-\frac{1}{2}\lambda' (P'^{^{t}} P' -1).
\end{equation}
where we will assume in the following that $\lambda,\lambda^\prime> 0 $. This problem can be solved using the $KKT$ conditions (see Appendix B). From condition $\nabla L=0$ , we have
\begin{equation}\label{op2}
\mathcal{A}P'=\lambda P ,\ \  P^{t}\mathcal{A}=\lambda' P'^{^{t}},
\end{equation}
and from the last KKT condition which we call complementary slackness, we get
\begin{equation}\label{op3}
\lambda (P^{t} P -1)=0  ,\ \  \lambda' (P'^{^{t}} P'-1)=0 ,
\end{equation}
which reduce to
$P^{t} P =1$ and $P'^{^{t}} P'=1$.
 Multiplying $\mathcal{A} P'=\lambda P$ by $P^{t}$ from the left side and by using $P^{t} P =1$, we obtain
$$
P^{t}\mathcal{A} P'=\lambda P^{t}P=\lambda,
$$
also multiplication $P^{t}\mathcal{A}=\lambda' P'^{^{t}}$ by $P'$ from the right side and $P'^{^{t}} P'=1$, yield
$$
P^{t}\mathcal{A} P'=\lambda' P'^{t}P'=\lambda',
$$
then $\lambda=\lambda'$. Therefore, from (\ref{op2}) and (\ref{op3}) one can write
$$
\mathcal{A}^{t}\mathcal{A} P'=\lambda^{2} P',
$$
and
$$
\mathcal{A} \mathcal{A}^{t}P=\lambda^{2} P,
$$
which indicate that $\lambda^{2}_{i}$ for $i=1,...,16$, are eigenvalues of $\mathcal{A}^{t}\mathcal{A}$ and $\mathcal{A}\mathcal{A}^{t}$. If we choose $\mathcal{A}$'s in a way that $\lambda^{2}_{i}\leqslant1$ (i.e. $-1\leqslant\lambda_{i}\leqslant1$) for all $i$'s then the minimum of Lagrangian (\ref{Lag1}) , i.e. $1+\lambda 1$ is nonnegative, which leads to the nonnegativity of $Tr(\mathcal{W}\rho_{_s})$.

{\bf The second step}

In the second step, for a given $\rho_{ent}$, we impose the second condition for an EW, $Tr(\mathcal{W} \rho_{ent})< 0$. Now the objective function ( which will be minimized ) is $Tr(\mathcal{W} \rho_{ent})$, and the inequality constraints come from the first step solution. So, this problem can be written as a convex optimization problem of the form
$$
\mathrm{Minimize} \quad Tr(\mathcal{W} \rho_{ent}),
$$
$$
s.t. \quad \mathcal{A}^{t}\mathcal{A}-I_{d}\leqslant 0,
$$
We can use associated Lagrange for solving this problem as
$$
L=Tr(\mathcal{W} \rho_{ent})+Tr[(\mathcal{A}^{t}\mathcal{A}-I_{d})Z],
$$
where $Z$ is a positive symmetric matrix. Then, using the $KKT$ (see Appendix B) conditions we get, after some mathematical manipulations, the following formula
$$
\tilde{\rho_1}+\mathcal{A}(Z^{t}+Z)=0,
$$
where $\tilde{\rho_1}$ is a matrix with components
\begin{equation}\label{RoTilta}
    \tilde{\rho_1}_{i,j}=Tr(\rho_1. \\ Q_{i} \otimes Q'_{j}),
\end{equation}
$$
\mathcal{A}^{t}\mathcal{A}=I_{d}.
$$
Obviously, these components are associated with components of witness matrix, $\mathcal{A}_{\mu,\nu}$.\\
If the matrix $Z$ be a symmetric matrix (i.e. $Z^{t}=Z$) then
$\tilde{\rho_1}=-2\mathcal{A}Z$ or
\begin{equation}\label{Z}
    Z=\frac{1}{2}\mathcal{A}^{t}\tilde{\rho_1}=\frac{1}{2}(\tilde{\rho_1} ^{t}\tilde{\rho_1})^{\frac{1}{2}},
\end{equation}
where
\begin{equation}\label{A}
    \mathcal{A}=-\frac{1}{2}\tilde{\rho_1} Z^{-1}.
\end{equation}
 Now the minimum of the Lagrangian becomes
\begin{equation}\label{MinLag}
    1-Tr[\sqrt{\tilde{\rho_1} ^{t}\tilde{\rho_1}} ],
\end{equation}
and the negativity of this term is the detection condition for the entanglement of the given density matrix.
The parameters of matrix $\mathcal{A}$ using the Eq.(\ref{A}) are calculated as
$$\mathcal{A}_{1,1}=\mathcal{A}_{2,2}=\mathcal{A}_{3,3}=\mathcal{A}_{4,4}=\frac{1}{2}(\frac{b_2-b_1}{|b_1-b_2|}-1),$$
$$\mathcal{A}_{1,2}=\mathcal{A}_{2,1}=\mathcal{A}_{3,4}=\mathcal{A}_{4,3}=\frac{1}{2}(\frac{b_1-b_2}{|b_1-b_2|}-1),$$
$$\mathcal{A}_{6,6}=\mathcal{A}_{9,9}=-\mathcal{A}_{12,12}=-\mathcal{A}_{15,15}=\frac{1}{2}(\frac{b_6-b_5}{|b_5-b_6|}-1),$$
$$\mathcal{A}_{6,9}=\mathcal{A}_{9,6}=-\mathcal{A}_{12,15}=-\mathcal{A}_{15,12}=\frac{1}{2}(\frac{b_5-b_6}{|b_5-b_6|}-1),$$
$$\mathcal{A}_{7,7}=\mathcal{A}_{8,8}=-\mathcal{A}_{13,13}=-\mathcal{A}_{14,14}=\frac{1}{2}(\frac{b_8-b_7}{|b_7-b_8|}-1),$$
$$\mathcal{A}_{7,8}=\mathcal{A}_{8,7}=-\mathcal{A}_{13,14}=-\mathcal{A}_{14,13}=\frac{1}{2}(\frac{b_7-b_8}{|b_7-b_8|}-1),$$
$$\mathcal{A}_{5,5}=\mathcal{A}_{10,10}=-1,$$
$$\mathcal{A}_{11,11}=\mathcal{A}_{16,16}=\left\{\begin{array}{c}-1 \ \ b_3 > b_4\\
+1 \ \ b_3 < b_4 \end{array}\right..$$
where
$$b_1= q_1+q_3+q_5+q_7,\quad
 b_2= q_9+q_{11}+q_{13}+q_{15},$$
 $$b_3= q_1-q_3-q_5+q_7,\quad
 b_4= q_9-q_{11}+q_{13}-q_{15},$$
 $$b_5= q_1-q_3+q_5-q_7,\quad
 b_6= q_9-q_{11}-q_{13}+q_{15},$$
 $$b_7= q_1+q_3-q_5-q_7,\quad
b_8= q_9-q_{11}-q_{13}-q_{15}.$$

As an example, let us consider the case  $b_1 > b_2,\  b_3 > b_4, \ b_5 > b_6$ and $ \ b_7 > b_8$ in which $\mathcal{A}_{i,j}$'s can be written as

$$\mathcal{A}_{i,j}=-1, \ \ i=j,$$
$$\mathcal{A}_{i,j}=0, \ \ i\neq j,$$
and it is easy to see that
$$\mathcal{W}=I_4\otimes I_4-(\mathcal{F}_1+\mathcal{F}_2+\mathcal{F}_3+\mathcal{F}_4),$$
where
$$\mathcal{F}_1=Q^{13}\otimes Q^{13}+Q^{14}\otimes Q^{14}+Q^{15}\otimes Q^{15}+Q^{16}\otimes Q^{16},$$
$$\mathcal{F}_2=Q^1_{S}\otimes Q^1_{S}+Q^6_{S}\otimes Q^6_{S}-Q^7_{A}\otimes Q^7_{A}-Q^{12}_{A}\otimes Q^{12}_{A},$$
$$\mathcal{F}_3=Q^2_{S}\otimes Q^2_{S}+Q^5_{S}\otimes Q^5_{S}-Q^{10}_{A}\otimes Q^{10}_{A}-Q^{9}_{A}\otimes Q^{9}_{A},$$
$$\mathcal{F}_4=Q^3_{S}\otimes Q^3_{S}+Q^4_{S}\otimes Q^4_{S}-Q^8_{A}\otimes Q^8_{A}-Q^{11}_{A}\otimes Q^{11}_{A}.$$
After some calculations one arrives at
\begin{equation}\label{tr1}
\mathcal{W}=I_4\otimes I_4-4\ket{\Phi^1}\bra{\Phi^1}.
\end{equation}
In the next section, we consider the optimality of DEW (\ref{tr1}) by using Hilbert-Schmidt measure and show that the DOEW (\ref{tr1}) is based on the minimal distance of an entangled state from edge of PPT states.

\section{Optimality DEW via Hilbert-Schmidt measure }
According to Ref.\cite{hilbersh}, the authors have used the Hilbert-Schmidt distance, which quantifies the distance of an
entangled state from the set of all separable states and calculated the optimal entanglement witnesses explicitly. However, in some situation, specially in high-dimensional Hilbert space we are not able to specify the region of separable states. In order to define this measure, we recall that the
Hilbert-Schmidt norm is defined as
\begin{equation}\label{HSn}
\|A\|=\sqrt{\langle A,A\rangle},
\end{equation}
where, $\langle A,B\rangle=Tr(A^{\dag}B)$. With help of the norm
(\ref{HSn}), the Hilbert-Schmidt distance between two arbitrary
states $\rho_A,\rho_B$ can be defined as
\begin{equation}\label{HSd}
d_{HS}(\rho_A,\rho_B)=\|\rho_A-\rho_B\|.
\end{equation}
By using the Hilbert-Schmidt distance, the so-called Hilbert-Schmidt
measure of entanglement is defined as
\begin{equation}\label{HS}
D(\rho_{ent.})=\min_{\rho\in S}\|\rho-\rho_{ent.}\|,
\end{equation}
where, $S$ is the set of separable states. In fact, the
Hilbert-Schmidt measure is the minimal distance of an entangled
state $\rho_{ent.}$ from the set of separable states.

For an entangled state $\rho_{ent}$, the minimum of the Hilbert-
Schmidt distance (the Hilbert-Schmidt measure) is attained for some
state $\rho_s$ since the norm is continuous and the set $S$ is
compact. Due to the Bertlmann-Narnhofer-Thirring Theorem \cite{16},
there exists an equivalence between the Hilbert-Schmidt measure and
the concept of optimal entanglement witnesses as follows: The
Hilbert-Schmidt measure of an entangled state equals the maximal
violation of the inequality $Tr(\mathcal{W}\rho)\geq 0$,
\begin{equation}\label{HSm}
D(\rho_{_{ent}})=\|\rho_s-\rho_{_{ent}}\|=-\langle\rho_{_{ent}},\mathcal{W}_{opt}\rangle=-Tr(\rho_{_{ent}}\mathcal{W}_{opt}),
\end{equation}
where,
\begin{equation}\label{HSm1}
\mathcal{W}_{opt}=\frac{\rho_s-\rho_{_{ent}}-\langle\rho_{s},\rho_{s}-\rho_{_{ent}}\rangle\mathbf{1}}{\|\rho_s-\rho_{ent.}\|},
\end{equation}
is an optimal entanglement witness (for more details see Refs.\cite{hilbersh,16}).\par

In this work we use this method in order to show the optimality of DEW (\ref{tr1}) and then we discuss the effects of Lorentz transformation on the measure of entanglement of two-particle states. One of the important problems is finding  nearest separable states to entangled state which is not easy for high-dimensional Hilbert space, so we consider the minimal distance of an entangled state from edge of PPT states and obtain optimal DEW.\\
For an entangled state $\rho_{ent}$, the minimum of the Hilbert-
Schmidt distance (the Hilbert-Schmidt measure) is attained for the edge of PPT
states, namely $\rho_{edge} .$ Then we replace $\rho_{s}$ with $\rho_{edge}$ in (\ref{HSm}) and (\ref{HSm1}) as
\begin{equation}\label{HSm2}
\mathcal{D}(\rho_{_{ent}})=\|\rho_{edge}-\rho_{_{ent}}\|=-\langle\rho_{_{ent}},\mathcal{W}_{Dopt}\rangle=-Tr(\rho_{_{ent}}\mathcal{W}_{Dopt}),
\end{equation}
and after some mathematical manipulations, we get
\begin{equation}\label{HSm2}
\mathcal{W}_{Dopt}=\frac{\rho_{edge}-\rho_{_{ent}}-\langle\rho_{edge},\rho_{edge}-\rho_{_{ent}}\rangle\mathbf{1}}{\|\rho_{edge}-\rho_{ent.}\|}=I_4\otimes I_4-4\ket{\Phi^1}\bra{\Phi^1},
\end{equation}\\
where $\rho_{edge}=\rho_{\frac{1}{4}}$ according to Eq.(\ref{fr1}) is the edge of PPT states and $\rho_{ent}=\rho_1$, so the DEW (\ref{tr1}) is optimal.

\section{Entanglement of relativistic two-particle quantum states}

Finally let us consider the amount of entanglement for relativistic pure and mixed states using the von Neumann entropy and DOEW, respectively.

\subsection{Entanglement of relativistic pure states}

Let $\ket{\psi}=\sum_{i,j,k,l=1}^2  a_{ijkl} \ket{p_i,j}\otimes \ket{p_k,l} , \ \   a_{ijkl}\in C$
be a general two-particle pure state
with normalization $\sum_{i,j,k,l=1}^{2}|a_{ijkl}|^2=1$. For this
pure state the entanglement of formation \emph{E} is defined as the
entropy of either of the two sub-systems, $\emph{E}(\ket{\psi})=-Tr(\sigma_1
Log_2 \sigma_1)=-Tr(\sigma_2 Log_2 \sigma_2),$ where
$\sigma_1$(respectively, $\sigma_2$) is the partial  trace of
$\ket{\psi}\bra{\psi}$ over the first (respectively, the second)
Hilbert space.
If $\mathcal{N}$
denotes the matrix with entries given by $a_{ijkl}$, then $\sigma_1$ can
be expressed as $\sigma_1=\mathcal{N}\mathcal{N}^\dagger $ ($\mathcal{N}^\dagger$ means conjugate transpose of $\mathcal{N}$).\par
In the previous section we considered two types of pure states which were different in combination of momentum eigen-state. If we calculate the von Neumann entropy for the pure states (\ref{iso}) under Lorentz transformation with arbitrary boost and momentum direction, we will see that these pure states does not change, except for the pure states of the first type  when  the momentum and  boost are parallel.
$$\emph{E}(\ket{\Lambda \Phi^{i}})=-\sum_{k}\lambda_k Log_2 \lambda_k,  \   \mathrm{\emph{i} = odd \ number,}$$
$$\lambda_{1(2)}=\frac{\cos^4({\frac{\theta_1}{2}})}{2(\cos^4({\frac{\theta_1}{2}})+\cos^4({\frac{\theta_2}{2}}))},$$
\begin{equation}\label{p11}
\lambda_{3(4)}=\frac{\cos^4({\frac{\theta_2}{2}})}{2(\cos^4({\frac{\theta_1}{2}})+\cos^4({\frac{\theta_2}{2}}))},
\end{equation}\\
where $\lambda_k$ is eigenvalues of the reduced density matrix ($\sigma_1$ or $\sigma_2$).
After some mathematical manipulations, we get
\begin{equation}\label{pure1}
\hspace{-5mm}\emph{E}(\ket{\Phi^i})=\frac{-1}{\cos^4({\frac{\theta_1}{2}})+\cos^4({\frac{\theta_2}{2}})}\{\cos^4({\frac{\theta_1}{2}})log(\frac{\cos^4({\frac{\theta_1}{2}})}{2(\cos^4({\frac{\theta_1}{2}})+\cos^4({\frac{\theta_2}{2}}))})+\cos^4({\frac{\theta_2}{2}})log(\frac{\cos^4({\frac{\theta_2}{2}})}{2(\cos^4({\frac{\theta_1}{2}})+\cos^4({\frac{\theta_2}{2}}))})\}.\\
\end{equation}\\
From this expression it can be appreciated that for $\theta_1=\theta_2$ when two particles in moving frame have the same velocity then the entanglement does not change but for $\theta_1\neq\theta_2$ the entropy decreases with increasing the velocities of the observers.

\subsection{Entanglement of relativistic mixed states using DOEW}

According to Eq.(\ref{MinLag}) and using the complete set of matrices $\{Q^i\}$  for $\rho_1$ , we get the following result:\\
\be\label{rpure}
\vspace{-10mm}\hspace{-30mm}Tr[\mathcal{W}\rho_1]=1-Tr[\sqrt{\tilde{\rho_1} ^{t}\tilde{\rho_1}} ]=\frac{1}{2}(1-|b_1-b_2|-|b_3-b_4|-|b_3+b_4|)$$

$$\hspace{50mm} -(|b_5-b_6|+|b_5+b_6|+|b_7-b_8|+|b_7+b_8|)\frac{\cos^2{\frac{\theta_1}{2}}\cos^2{\frac{\theta_2}{2}}}{\cos^4{\frac{\theta_1}{2}}+\cos^4{\frac{\theta_2}{2}}},\ee\\
This result shows that the entanglement  between two particles which are described by the mixed density matrix $\rho_1$, under Lorentz transformations is not Lorentz-invariant and decreases by increasing the velocity of the observable.\\
For detection of entanglement of the relativistic mixed density matrix $\rho_1$  using DOEW (\ref{tr1}) , the following result is obtained
\begin{equation}\label{tr2}
Tr[\mathcal{W}.\rho_1]=1-2q_1-2q_7+4(q_7-q_1)\frac{\cos^2{\frac{\theta_1}{2}}\cos^2{\frac{\theta_2}{2}}}{\cos^4{\frac{\theta_1}{2}}+\cos^4{\frac{\theta_2}{2}}}.
\end{equation}\\
Therefore, we have
\begin{equation}\label{tr3}
Tr[ \mathcal{W}.\rho_1]\geq Tr[ \mathcal{W}.\rho_1]_{rest}.
\end{equation}\\
This result indicates that when momentum and boost are parallel then the entanglement is not Lorentz-invariant because $Tr[ \mathcal{W}.\rho_1]_{rest}$ is more negative than (\ref{tr2}). Therefore, these DOEWs can be used to quantify
the amount of entanglement. At the end of this section we show that the result (\ref{tr2}) can be associated with generalized concurrence and coincides  with previous result for pure state in (\ref{pure1}).\\
According to Ref.\cite{Shao} the entanglement of formation of $\ket{\Phi^i}$ is given by
$$\emph{E}(\ket{\Phi^1})=-n\lambda_1 Log_2 \lambda_1-m\lambda_2 Log_2 \lambda_2$$
where $\lambda_{1,2}$ are defined in (\ref{p11}) and  $ n=m=2 , \lambda_1+\lambda_2=\frac{1}{2}.$ Using the expectation value (\ref{tr3}) for density matrix $\ket{\Phi^1}\bra{\Phi^1}$  we have
$$Tr[\mathcal{W}.\rho_1]=-1-4\frac{\cos^2{\frac{\theta_1}{2}}\cos^2{\frac{\theta_2}{2}}}{\cos^4{\frac{\theta_1}{2}}+\cos^4{\frac{\theta_2}{2}}}= -1-8\sqrt{\lambda_1\lambda_2},$$\\
where $q_i$'s are zero except for $q_1=1$. That is, if we define $\chi=1+8\sqrt{\lambda_1\lambda_2}$, then

$$\lambda_1=\frac{1}{2}\{\frac{1}{2}+\sqrt{\frac{1}{4}(1-\frac{(\chi+1)^2}{4})}\}=\frac{1}{2}\{\frac{1}{2}+\sqrt{1-d^2}\},$$

$$\lambda_2=\frac{1}{2}\{\frac{1}{2}-\sqrt{\frac{1}{4}(1-\frac{(\chi+1)^2}{4})}\}=\frac{1}{2}\{\frac{1}{2}-\sqrt{1-d^2}\},$$\\
where \emph{d} is generalized concurrence which has been defined as $\emph{d}=4\sqrt{\lambda_1\lambda_2}$.


\section{Conclusions}

We have obtained  DOEWs by using the convex optimization method and  shown that it is possible to use DOEW operators to quantify the amount of entanglement of bipartite mixed states for high-dimensional Hilbert space. This was due to the fact that, one can calculate the minimum distance of an entangled state from the edge of bound entangled states, via the optimality DEWs. Then, by using it, we have discussed the entanglement of quantum states of two $s=\frac{1}{2}$ fermions  under an arbitrary Lorentz transformation. For convenience, instead of
using the superposition of momenta we used only two momentum
eigen states ($p_1$ and $p_2$). Consequently, in 2D momentum subspace we have shown that for relativistic mixed density matrix $\rho_1$ when momentum and boost are parallel the entanglement is not Lorentz-invariant.\\
Our method can be generalized for applying  the techniques of this
paper to  quantifying the amount of entanglement of bound entangled states that we can evaluate the minimum distance of  bound entangle state from the edge of PPT states but in the orientation of inside to separable states direction.


  \vspace{1cm}\setcounter{section}{0} \setcounter{equation}{0}
\renewcommand{\theequation}{A-\roman{equation}}
{\Large APPENDIX A}

{\bf Wigner representation for spin-$\frac{1}{2}$:}

It follows \cite{weinberg} that the effect of an arbitrary Lorentz
transformation $\Lambda$ unitarily implemented as $U(\Lambda)$ on
single-particle states is as follow \be
\label{wig1}U(\Lambda)(\ket{p}\otimes\ket{\sigma})=\sqrt{\frac{(\Lambda
p)^0}{p^0}}\sum_{\sigma^\prime} D_{\sigma^\prime
\sigma}(W(\Lambda,p))(\ket{\Lambda p}\otimes\ket{\sigma^\prime})
,\ee where \be \label{wig11}W(\Lambda,p)=L^{-1}(\Lambda p)\Lambda
L(p),\ee\\ is the Wigner rotation \cite{wigner} .The Wigner rotation
is an element of the spacial rotation group $SO(3)$ or subgroup of
the homogeneous Lorentz group since it leaves the rest momentum $
k^\nu$ unchanged :
$$W_{\nu}^\mu k^\nu=k^\mu .$$
This subgroup is called (Wigner's) little group. We will consider two
reference frames in this work: one is the rest frame S and the other
is the moving frame $S^\prime$ in which a particle whose
four-momentum \emph{p} in S is seen as boosted with the velocity
$\vec{v}$. By setting the boost and particle moving directions in
the rest frame to be $\hat{v}$ with $\hat{e}$ as the normal vector
in the boost direction  and $\hat{p_{1(2)}}$, respectively, and
$\hat{n}=\hat{e}\times \hat{p_{1(2)}}$, the Wigner representation for
spin-1/2 is found as \cite{Ahn},
\be
\label{wig2}D^{\frac{1}{2}}(W(\Lambda,p_{1(2)}))=\cos{\frac{\Omega_{\vec{p_{1(2)}}}}{2}}+i\sin{\frac{\Omega_{\vec{p_{1(2)}}}}{2}}(\vec{\sigma}.\hat{n}),\ee\\
where \be
\label{wig3}\cos{\frac{\Omega_{\vec{p_{1(2)}}}}{2}}=\frac{\cosh{\frac{\alpha}{2}}\cosh{\frac{\delta}{2}}+\sinh{\frac{\alpha}{2}}\sinh{\frac{\delta}{2}}(\hat{e}.\hat{p_{1(2)}})}{\sqrt{[\frac{1}{2}+\frac{1}{2}\cosh{\alpha}\cosh{\delta}+\frac{1}{2}\sinh{\alpha}\sinh{\delta}(\hat{e}.\hat{p_{1(2)}})]}},\ee\\
\be
\label{wig4}\sin{\frac{\Omega_{\vec{p_{1(2)}}}}{2}}\hat{n}=\frac{\sinh{\frac{\alpha}{2}}\sinh{\frac{\delta}{2}}(\hat{e}\times\hat{p_{1(2)}})}{\sqrt{[\frac{1}{2}+\frac{1}{2}\cosh{\alpha}\cosh{\delta}+\frac{1}{2}\sinh{\alpha}\sinh{\delta}(\hat{e}.\hat{p_{1(2)}})]}},\ee\\
$$(\cos{\frac{\Omega_{\vec{p_{1(2)}}}}{2}})^2+(\sin{\frac{\Omega_{\vec{p_{1(2)}}}}{2}}\hat{n})^2=1,$$
 and
$$\cosh{\alpha}=\gamma=\frac{1}{\sqrt{1-\beta^2}} ,\cosh{\delta}=\frac{\emph{E}}{m} ,\beta=\frac{v}{c}
.$$\\

{\Large APPENDIX B}

{\bf Convex optimization and Karush-Kuhn-Tucker (KKT) Theorem:}

Many problems from quantum information theory can easily be translated to the language of convex optimization problems, of the form

\hspace{-40mm}\begin{equation}\hspace{-30mm}\label{COP}
 \begin{array}{c}
  \mathrm{minimize} \      \ \hspace{10mm}f_{0}(x) \\
   \hspace{32mm}\mathrm{subject} \ \mathrm{to} \hspace{8mm}\ f_{i}(x)\leqslant 0$,   \hspace{5mm}$i=1,...,m. \\
   \hspace{57mm}h_{j}(x)= 0$,   \hspace{5mm}$j=1,...,p. \\
 \end{array}
\end{equation}
where x is a vector of decision variables, and the
functions $f_0$ , $f_i$ and $h_j$ , respectively, are the cost, inequality and equality constrains which satisfy inequality $f_{i}(\alpha x +\beta y)\leqslant \alpha f_{i}(x)+\beta f_{i}(y)$, for all $x, y \in R$ and all $\alpha, \beta \in R$ with $\alpha +\beta = 1$, $\alpha \geqslant 0$, $\beta \geqslant 0$ because of convexity and the equality constraint functions $h_{i}(x)= 0$ must be affine (A set $C\in\textbf{R}^{n}$ is affine if the line through any two distinct points in $C$ lies in $C$).\\
There are a number of important necessary conditions that hold for problems with zero
duality gap. These Karush-Kuhn-Tucker conditions turn out to be sufficient for convex
optimization problems. Assuming that functions $f_i$ and $h_j$  are differentiable and that strong duality holds, there exists vectors $\zeta\in R^k$,  and $y\in R^m$, such that the gradient of dual Lagrangian $L(x^\ast,\zeta^\ast,y^\ast)=f(x^\ast)+\sum_i \zeta_i^\ast h_i(x^\ast)+ \sum_i y_i^\ast g_i(x^\ast)$ over x vanishes at $x^\ast$

$$h_i(x^\ast)=0 \ \  (primal feasible)$$
$$g_i(x^\ast)\leq0 \  \ (primal feasible)$$
$$y_i^\ast\geq0 \  \  (dualfeasible)$$
$$y_i^\ast g_i(x^\ast)=0$$
$$\nabla f(x^\ast)+\sum_i \zeta_i^\ast \nabla h_i(x^\ast)+ \sum_i y_i^\ast \nabla g_i(x^\ast)=0.$$\\
Then $x^\ast$ and $(\zeta^\ast ,y^\ast)$ are primal and dual optimal, with zero duality gap. In summary, for any convex optimization problem with differentiable objective and constraint functions, any points that satisfy the KKT conditions are primal and dual optimal, and have zero duality gap.
Necessary KKT conditions satisfied by any primal and dual optimal pair and for convex problems, KKT conditions are also sufficient (see
Ref.\cite{Boyd} for more details) .

\end{document}